\documentclass[aps,pra,twocolumn,superscriptaddress,amsmath,amssymb,revsymb,showpacs]{revtex4}

\usepackage{graphicx}
\usepackage{dcolumn}
\usepackage{bm}
\usepackage[usenames]{color}
\newcommand{\be}{\begin{eqnarray}}
\newcommand{\ee}{\end{eqnarray}}

\newcommand{\nn}{\nonumber}

\newcommand{\bo}{\boldsymbol}
\newcommand{\lb}{\label}
\begin{document}


\title{Emergence of $\eta$-pairing ground-state in population-imbalanced attractive Fermi-gases filling $p$ orbitals on 1-D optical lattice }

\author{Keita Kobayashi}%
\affiliation{%
CCSE, Japan Atomic Energy Agency, Kashiwa, Chiba
 277-0871, Japan}%
\author{Yukihiro Ota}%
\affiliation{%
CCSE, Japan Atomic Energy Agency, Kashiwa, Chiba
 277-0871, Japan}%
\author{Masahiko Okumura}%
\affiliation{%
CCSE, Japan Atomic Energy Agency, Kashiwa, Chiba
 277-0871, Japan}%
\author{Susumu Yamada}%
\affiliation{%
CCSE, Japan Atomic Energy Agency, Kashiwa, Chiba
 277-0871, Japan}%
\affiliation{%
Computational Materials Science Research Team, RIKEN AICS, Kobe, Hyogo
650-0047, Japan 
}%
\author{Masahiko Machida}%
\affiliation{%
CCSE, Japan Atomic Energy Agency, Kashiwa, Chiba
 277-0871, Japan}%
\date{\today}

\begin{abstract}

We explore the ground states in population-imbalanced attractive 1-D
fermionic optical lattice filling $p$ orbitals over the lowest $s$ one by using the
density-matrix-renormalization-group (DMRG) method. 
The DMRG calculations find the occurrence of spatially non-uniform 
off-diagonal long-range order. 
In contrast to Fulde-Ferrel Larkin-Ovchinikov pair 
as observed in the single-band Hubbard model. 
The spatial oscillation period of 
the pair correlation function is widely fixed to be $\pi$ irrespective of the mismatch 
between spin-split Fermi surfaces. 
The ground-state $\pi$ order corresponds to $\eta$-pair condensate predicted
by Yang [Phys. Rev. Lett. \textbf{63}, 2144 (1989)]. 
Taking account of the effects of harmonic traps, we confirm that
the $\eta$-pair state distinctly emerges at the center of the trap potential surrounded 
by perfectly-polarized states even in the trapped cases. 

\end{abstract}

\maketitle

\section{Introduction}

Ultracold atomic gases offer excellent playgrounds to study many-body quantum effects.  
Indeed, atomic Fermi gases loaded on optical lattices have 
attracted much attention as virtual strongly-correlated electron systems~\cite{Bloch:Zwerger:2008,Bloch:Nascimnene:2012}.  
One of the advantages using optical lattices is the tunability of interaction which ranges 
from attractive to repulsive through the Feshbach resonance. 
Moreover, various kinds of lattice configurations are available. 
The uniquness opens a field studing not only unsolved issues in 
condensed matter physics like magnetism and High-$T_{\rm c}$
superconductivity but also imaginary exotic ordered states. 

Manipulating orbital degrees of freedom in atomic gases is presently of particular
interest~\cite{Muller;Bloch:2007,Wu;Zhai:2008,Wu:2008,Wirth;Hemmerich:2011,Kobayashi;Machida:2012,Kobayashi;Machida:2014}.  
The marked effect of orbitals is to provide rich internal degrees of
freedom into atomic gases in addition to spin one. 
The multiple orbital degrees of freedom lead to 
non-trivial structures of quantum phases like the Haldane phase as seen in fermionic 1D
chains~\cite{Kobayashi;Machida:2012,Kobayashi;Machida:2014,Nonne;Boulat:2010,Nonne;Totsuka:2014}. 
To reveal which types of quantum phases the orbital degrees of freedom bring is 
an attractive issue in exploring intriguing varieties of many-body quantum phenomena. 

In this article, we study interplay of orbital degrees of freedom 
and population imbalance in an attractively-interacting Fermi 1D chain
with double degenerate $p$-orbital using the
density matrix renormalization group (DMRG) method~\cite{White:1992,Schollwock:2011,Yamada;Machida:2009}. 
The population imbalance in attractive Fermi gases allows to observe the Fulde-Ferrell~\cite{Fulde;Ferrell:1964} Larkin-Ovchinnikov~\cite{Larkin;Ovchinnikov:1965} (FFLO) superfluid
phase (See, e.g., Refs.~\cite{Partridge;Hulet:2006,Feiguin;HeidrichMeisner:2007,Tezuka;Ueda:2008,Feiguin;HeidrichMeisner:2009,Liao;Mueller:2010,
Zi;Cai:2011,Lu;Pu:2012,Swanson;Trivedi:2012}). 
The Cooper pair in FFLO phase has non-zero center-of-mass momentum, and then the order parameter is 
non-uniform in the real-space. 
Our calculations confirm the emergence of FFLO phase in the present $p$-orbital model with power
law decay, while a more striking result is the occurrence of 
another quantum phase leading
to the long-range correlation with non-zero center-of-mass momentum. 
This phase corresponds to the off-diagonal-long-range ordered state predicted by
Yang~\cite{Yang:1989} called as $\eta$-pair condensate state. 
The pairing mechanism of the FFLO state is responsible for the spin-split Fermi
surfaces, i.e., the center-of-mass momentum in the pair depends on the
mismatch of the Fermi surfaces. 
In contrast, the center-of-mass momentum of $\eta$-pair is always fixed as
$\pi$ owing to pseudo-spin $SU(2)$ symmetry in the Hubbard 
Hamiltonian~\cite{Yang:1989,Demler;Scalapino:1996}. 
The possibility of $\eta$-pair has been recently discussed in the context of Iron-based 
superconductors without spin-imbalance
~\cite{Hu;Hao:2012,Gao;Zhu:2010,Khodas;Ghubukov:2012,Hu;2013} 
and fermionic system on optical lattices with ac fields~\cite{Kitamura_Aoki;2016}.
We clarify how $\eta$-pairing phases emerge in the present system by 
examining the pair correlation function based on DMRG calculations.

We stress that the occurrence of $\eta$-pairing is notable, since
this state is an eigenvector but not the ground state 
of the single-orbital Hubbard Hamiltonian~\cite{Yang:1989,Demler;Scalapino:1996}. 
The underlying mechanism of the $\eta$-pairing in the present system is
revealed, in terms of the repulsive-attractive
transformation~\cite{Shiba:1972,Micnas;Robaszkiewicz:1990}. 
In addition, we study the effect of the harmonic trap potential on the
formation of the $\eta$-pairing state, towards the experimental
verification. 
We show that the $\eta$-paring state in a trapped system occurs at the
center of the trap potential, surrounded by a spin-polarized state.



The paper is organized as follows. 
The Hamiltonian for a uniform 1D $p$-orbital chain is shown in
Sec.~\ref{sec:model}, with a brief summary of the properties without the
population imbalance. 
Section \ref{sec:results} is the main part of this article. 
The DMRG-calculation results are shown for the uniform system
(Sec.~\ref{subsec:uniform}) and the trapped system
(Sec.~\ref{subsec:trap}), focusing on the pair correlation function. 
At the end of Sec.~\ref{subsec:uniform}, the occurrence of the
$\eta$ pairing state is explained, in terms of the repulsive-attractive
transformation. 
We summarize the results in Sec.~\ref{sec:summary}. 

\section{Model}
\label{sec:model}
We study 1D Fermi gases with partially-filled $p$-orbital and 
completely-filled lower-energy orbital~\cite{Kobayashi;Machida:2012,Kobayashi;Machida:2014} (see also appendix A). 
This model is attainable in two-component Fermi gases (atomic
gases with spin degrees) on an optical lattice highly elongated
along one direction (i.e., $z$-axis). 
The cylindrically-symmetric potential perpendicular to
$z$-axis (i.e., potential on $xy$-plane) induces orbital characters into
the system. 
The spatial symmetry indicates that the energy levels with respect to
$p$-orbital are degenerate. 
The double degenerate $p_{x}$ and $p_{y}$ orbitals are active degrees of
freedom. 
The model Hamiltonian with total spatial sites $L$ is given by
\begin{eqnarray}
H
&=&
\sum_{i=1}^{L}
\bigg[
-t
\sum_{p,\sigma}
(
c_{p,\sigma,i+1}^{\dagger}c_{p,\sigma,i}
+
\text{h.c.}
) \nn \\
&&
-\mu 
\sum_{p,\sigma}
n_{p,\sigma,i} 
-h
\sum_{p}
S_{p,i}^{z} 
\nonumber \\
&&
+
\sum_{p}
U_{pp}
\left(n_{p,\uparrow,i}-\frac{1}{2}\right)
\left(n_{p,\downarrow,i}-\frac{1}{2}\right)
\nonumber \\
&&
+
\sum_{p\neq p^{\prime}}
U_{pp'}
\Big{ ( }
\bo{\rho}_{p,i}\cdot\bo{\rho}_{p',i}
-\bo{S}_{p,i}\cdot\bo{S}_{p',i}
\Big{ ) }
\bigg] .
\label{eq:p1d}
\end{eqnarray}
where 
$n_{p,\sigma,i} (= c^{\dagger}_{p,\sigma,i}c_{p,\sigma,i})$ is the on-site number 
operator for the $p$-orbital with spin $\sigma$. The spin-$1/2$ and 
the pseudo-spin-$1/2$ operators are, respectively, 
\mbox{
\(
\bo{S}_{p,i}
=\frac{1}{2}\bo{c}_{p,i}^\dagger\bo{\tau}\bo{c}_{p,i}
\)}
and 
\mbox{\(
\bo{\rho}_{p,i}
=\frac{1}{2}\bo{\tilde{c}}_{p,i}^\dagger\bo{\tau}\bo{\tilde{c}}_{p,i}
\)}, 
with
\mbox{
$\bo{c}_{p,i}=\!^{\rm t}(c_{p,\uparrow,i},c_{p,\downarrow,i})$} 
and 
\mbox{\(
\bo{\tilde{c}}_{p,i}
=\!^{\rm t}(c_{p,\uparrow,i},c_{p,\downarrow,i}^\dagger )
\)}, 
where 
\mbox{$\bo{\tau}=(\tau^{x},\tau^{y},\tau^{z})$} are the $2\times 2$ Pauli
matrices.  
The physical variables in Eq.~(\ref{eq:p1d}) are the hopping matrix 
element $t$ along the chain, the chemical potential $\mu$, the
magnetic-field strength $h$, and intra-/inter-orbital on-site coupling
constants $U_{pp^{\prime}}$. 
The orbital degeneracy leads to $U_{p_{x}p_{x}}=U_{p_{y}p_{y}}$, and 
then the intra-orbital coupling constant is denoted as $U_{pp}$. 
We always set attractive two-body interaction, i.e., $U_{pp}<0$ through this article. 
In typical atomic-gas experiments, the introduction of 
the population imbalance changes $h$. 

Here, we briefly summarize the properties of this
system, without the population imbalance ($h=0$). 
When balanced ($h=0$), we have 
\mbox{$SU(2)_{\rm spin}\times SU(2)_{\text{pseudo-spin}}$}  
symmetry of Eq.~(\ref{eq:p1d}) with the half filling ($\mu=0$). 
By taking the strong coupling limit ($|U_{pp}| \gg t $), this system is reduced to be the
pseudo-spin-$1$ Heisenberg model~\cite{Kobayashi;Machida:2014},  
\be
H_{\rm ps}=\sum_{<i,j>}J_{\rm ex}(\rho_{i}^{z}\rho_{j}^{z}-\rho_{i}^{x}\rho_{j}^{x}-\rho_{i}^{y}\rho_{j}^{y})
-2\mu\sum_{i}\rho_{i}^{z}\,,
\ee
with $\bo{\rho}_{i}=\sum_{p}\bo{\rho}_{p,i}$ and 
$J_{\rm ex}=2t^2/(|U_{pp}|+|U_{p_xp_y}|)$.  
According to the Haldane conjecture~\cite{Haldane_s}, a charge gap
(pseudo-spin gap) opens at half filling ($\mu=0$), 
where density and pair correlation functions 
(longitudinal and transverse correlation of pseudo spins) decay 
exponentially. 
This is a Haldane insulator phase. 
The formation of the Haldane insulator phase is proposed in different
atomic-gas systems, such as bosonic~\cite{DallaTorre_s} 
and fermionic
chains~\cite{Nonne;Totsuka:2014,Kobayashi;Machida:2014}. 
Below the half filling, the charge gap closes.  
The spin chain is regarded as the Luttinger liquids~\cite{LL1_s}; both the
charge and the pair correlations decay, with a power law. 
To sum up, when the population imbalance is absent, the properties of
the $p$-orbital 1D chain is similar to a simple 1D chain, except the
half-filling case.

\section{Numerical Results}
\label{sec:results}

We examine the ground-states in the $p$-orbital 1D chain
(\ref{eq:p1d}) under non-zero population imbalances by using the 
density-matrix-renormalization-group (DMRG) 
method~\cite{White:1992,Schollwock:2011,Yamada;Machida:2009}. 
The number of the states kept is varied from $400$ to maximally
$1500$ depending on the convergence tendency of the calculations.  
Imposing the open boundary conditions, we perform 
the calculations with varying the intra-orbital coupling, the
filling, and the population imbalance. 
The inter-orbital coupling is fixed by a relation 
\(
U_{p_{x}p_{y}} = U_{pp}/3
\)~\cite{Kobayashi;Machida:2012}. 

To achieve the
variations of the filling and the population imbalance 
instead of $\mu$ and $h$ in Eq.~(\ref{eq:p1d}), we change
$N_{\uparrow} + N_{\downarrow }$ and $N_{\uparrow} - N_{\downarrow}$ as
input parameters, where the spin-up particle number is $N_{\uparrow}$
and the spin-down $N_{\downarrow}$. 
In the uniform system, we use the filling rate per orbital and the
polarization rate per orbital, respectively, defined as
$\tilde{n} = (N_{\uparrow} + N_{\downarrow})/4L$ 
and 
$\tilde{m} = (N_{\uparrow} - N_{\downarrow})/2L$, 
since the intensive quantities well characterize system features.


\begin{figure}[h]
\begin{center}
\includegraphics[width=1.0\linewidth]{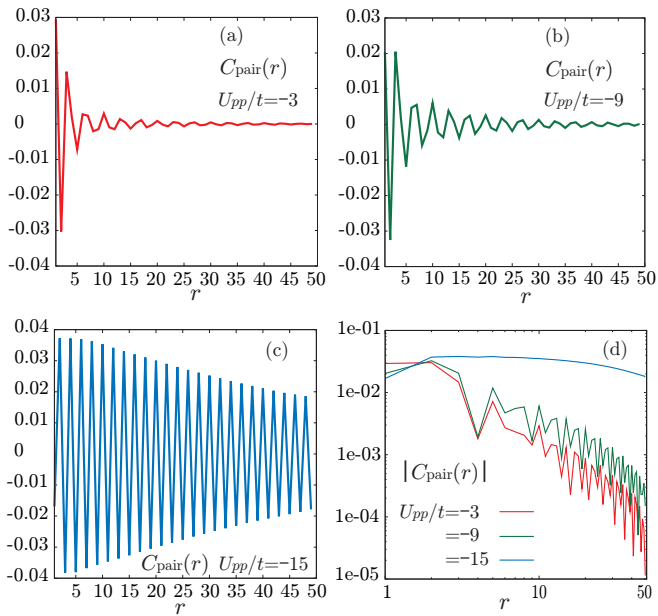}
\end{center}
\caption{(Color online) Spatially averaged pair correlation function
 $C_{\rm pair}(r)$ as a funtion of relative site difference $r$, for on-site
 intra-orbital interaction strengths (a) $U_{pp}/t=-3$, (b) $U_{pp}/t=-9$, and
 (c) $U_{pp}/t=-15$, in uniform $p$-orbital 1D chain (\ref{eq:p1d}), respectively.  
The absolute value for each $U_{pp}/t$ in a log-log plot is shown in (d). 
The total number of the sites is $L=100$, and the filling rate and the magnetization 
density rate are, respectively,  
$\tilde{n}=0.88$ and $\tilde{m}=0.6$ (i.e., $N_{\uparrow}=148$ and
 $N_{\downarrow}=28$). 
}
\label{Fig1}
\end{figure}

\subsection{Uniform System}
\label{subsec:uniform}

First, we calculate the spatial correlation function for
the on-site singlet pair, 
$\Delta_{p,i}=c_{p,\downarrow,i}c_{p,\uparrow,i}$. 
The two-point pair correlation function between sites $i+r$ and $i$ 
$
 C_{\rm pair}(i+r,i)=\langle
\Delta_{p,i+r}
\Delta_{p,i}^\dagger
\rangle,
\label{eq:pair_correlation}
$ 
where $\langle\cdot\rangle$ is the expectation values in the ground state of
Eq.~(\ref{eq:p1d}). 
A way to minimize finite-size effects is required to check
how long the correlation spatially extends. 
We use the spatial averaged pair correlation function 
\mbox{
$
C_{\rm pair}(r)
=[1/(L-2i_0-r)]
\sum_{i=i_{0}+1}^{L-i_{0}-r}C_{\rm pair}(i+r,i) 
$}~\cite{Alejandro}, where 
the summation is taken over an inner domain inside the system chain, 
\([i_{0}+1,\, L-i_{0}-r]\) with an edge cut-off parameter $i_{0}$. 
Considering the uniform system, we focus on the case below the half
filling ($\tilde{n}<1$) with positive imbalance ratio 
($\tilde{m} >0$). 
We note that the results for $\tilde{n}>1$ and $\tilde{m}<0$ are identical 
with the ones for $2-\tilde{n}$ and $-\tilde{m}$\,. 

Figure \ref{Fig1} shows $C_{\rm pair}(r)$ for different
values of $U_{pp}/t$.  
The filling and polarization rates are fixed as $\tilde{n}=0.88$
and $\tilde{m}=0.6$. 
The total number of the spatial sites is $L=100$, and the positional 
number of the edge cut-off sites is $i_{0}=25$. 
We find a damped oscillating behavior for $U_{pp}/t=-3$ and $-9$ 
[Fig.~\ref{Fig1}(a) and (b)].    
In contrast, for strong attractive interaction $U_{pp}/t=-15$, a
staggered oscillating behavior is observed [Fig.~\ref{Fig1}(c)]. 
Figure \ref{Fig1}(d) shows the decaying manner of 
$C_{\rm pair}(r)$ more clearly.  
The power-law decay is found for $U_{pp}/t=-3$ and $-9$, 
whereas the decay for $U_{pp}/t=-15$ is much slower.

\begin{figure}[h]
\begin{center}
\includegraphics[width=1.0\linewidth]{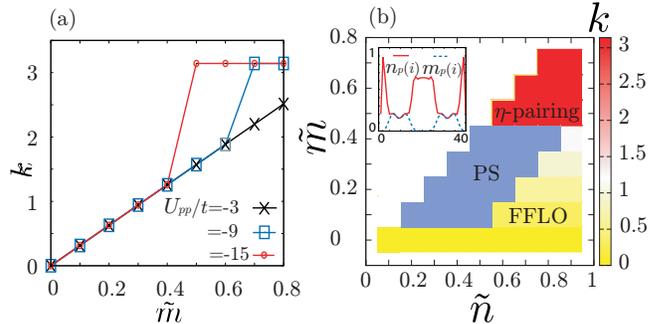}
\end{center}
\caption{(Color online) 
(a) Wave vector (spatial periodicity) $\bar{k}$
 of a two-point pair correlation function 
 versus the magnetization rate $\tilde{m}$, for the intra-orbital
 interaction strengths $U_{pp}/t=-3$, $-9$, and $-15$, in uniform
 $p$-orbital 1D chain (\ref{eq:p1d}). 
The filling rate is $\tilde{n}=0.875$. 
The total number of the sites is $L=40$. 
(b) Phase diagram of the ground states in Eq.~(\ref{eq:p1d}),
 for $U_{pp}/t=-15$.  
The Fulde-Ferrell and Larkin-Ovchinnikov superfluid phase is
 denoted by FFLO. 
The phase-separation phase is by PS. 
The $\eta$-pairing phase is by $\eta$-pairing.  
The inset shows the particle-density profile $n_{p}(i)=2\langle \rho^{z}_{p,i}\rangle +1$ and the
 spin-density profile $m_{p}(i)=2\langle S^{z}_{p,i}\rangle$, to
 show the phase-separated behavior in PS. 
}
\label{Fig2}
\end{figure}

Now, let us classify the quantum phases below the half-filling in non-zero
imbalance ratios.  
We focus on the spatial period of the two-point pair correlation
function $C_{\rm pair}(i_{\rm c}+r,\,i_{\rm c})$ in the central site
of the chain, $i_{\rm c}=L/2$.  
Performing the Fourier transformation of 
$C_{\rm pair}(i_{\rm c}+r,\,i_{\rm c})$ with respect to $r$, we obtain
the wave vector $\bar{k}$ characterizing the spatial period. 
Figure \ref{Fig2}(a) shows the wave vectors $\bar{k}$ for 
different $\tilde{m}$ and $U_{pp}/t$. 
The filling is fixed as $\tilde{n} = 0.875$, which 
is almost the same as in
Fig.~\ref{Fig1}. 
Figure \ref{Fig2}(b) shows $\bar{k}$ on 
\mbox{$\tilde{n}$-$\tilde{m}$} plane at $U_{pp}/t=-15$ (strong
inter-orbital interaction) for $L=40$. 
We find that the quantum phases are classified into three classes.  
When giving a small imbalance rate [e.g., $\tilde{m}<0.4$ for
$U_{pp}/t=-15$ in Fig.~\ref{Fig2}(a)], the wave vector
$\bar{k}$ is proportional to $\tilde{m}$, i.e., $\bar{k}=\pi\tilde{m}$.  
This phase is considered to be the FFLO superfluid phase, since
$\bar{k}$ is equal to the difference between the $p$-orbital Fermi wave
vectors, $k_{{\rm F},p,\uparrow}-k_{{\rm F},p,\downarrow}$ 
with 
$
k_{{\rm F},p,\uparrow(\downarrow)}=
\sum_{i}\pi n_{p,\uparrow(\downarrow),i}/L
$. 
The second phase is a phase-separated (PS) phase for small $\tilde{n}$
and large $U_{pp}/t$ as seen in Fig.~\ref{Fig2}(b) (the
gray-colored region).  
In the inset of Fig.~\ref{Fig2}(b), we exhibit the emergence of
the phase-separated polarized regions. 
The third phase occurs when both $\tilde{m}$ and $U_{pp}/t$ are large. 
Then, the wave vector $\bar{k}$ is just equal to $\pi$ independent of the value of $\tilde{m}$. 
Since the pairing state has always center-of-mass momentum equal to $\pi$, 
it is clear that an $\eta$-pairing state~\cite{Yang:1989} emerges in this range
($\eta$-pairing phase). 

Now, let us study the origin of the $\eta$-paring state. 
For the purpose, we apply attractive-repulsive
transformation~\cite{Shiba:1972,Micnas;Robaszkiewicz:1990},
$
c_{p,\uparrow,j}\to\bar{c}_{p,\uparrow,j}
$ and 
$
c_{p,\downarrow,j}^\dagger\to e^{i\pi x_{j}}\bar{c}_{p,\downarrow,j}
$
into Eq.~(\ref{eq:p1d}), where $x_{j}$ is the coordinate of the site $j$
along the chain. 
Writing the parameter- and the operator-dependences explicitly,  
\(
H = H
(
\mu,\,h,\, U_{pp^{\prime}},\, c_{p,\sigma,i},\, c^{\dagger}_{p,\sigma,i}
)
\), 
we find that the transformed Hamiltonian  
$
\bar{H}
=
H(
h,\,\mu,\, -U_{pp^{\prime}},\, \bar{c}_{p,\sigma,i},\, 
\bar{c}^{\dagger}_{p,\sigma,i}
)
$\,, 
where the attractive interaction $U_{pp'}$ becomes the repulsive one, the role of the chemical potential $\mu$ (filling) is replaced 
by the magnetic field $h$ (population imbalance), and vice
versa. 
The pair correlation function 
$\langle
\Delta_{p,i+r}
\Delta_{p,i}^\dagger
\rangle$ is transformed into the magnetic correlation for transverse direction 
$\langle\bar{S}_{p,i+r}^{-}\bar{S}_{p,i}^{+}\rangle$, 
where $\bo{\bar{S}}_{p,i}$ is the transformed spin-$1/2$ operator. 
Then, the FFLO state correspond to the transverse component of the 
spin density wave whose spatial modulation is characterized by the density of holes. 
We stress in the transformed system that the particles on $p_x$- and
$p_y$-orbitals interact with each other via Hund's coupling, 
$
-|U_{pp'}|\bo{\bar{S}}_{p,j}\cdot\bo{\bar{S}}_{p',i}
$ 
and a hole-doped two-degenerate band system 
with the Hund's coupling exhibits the 
ferromagnetism in the strong coupling 
range~\cite{Kusakabe;Aoki:1994,Momoi;Kubo:1998}. 
Moreover, it is remarkable that the long-range order of the ferromagnetic correlation for the
transverse direction, i.e., 
$
\lim_{r\to\infty}\langle\bar{S}_{p,i+r}^{-}\bar{S}_{p,i}^{+}\rangle
={\rm const,}
$ 
transforms into the pair correlation function with $\pi$ phase variation, 
$
\lim_{r\to\infty}\langle
\Delta_{p,i+r}
\Delta_{p,i}^\dagger
\rangle
=e^{-i\pi r}\times{\rm const}
$.
Consequently, one finds that 
the $\eta$-pairing state corresponds to the ferromagnetic ground-state in the repulsive model through 
the transformation.
Namely, the growth of the off-diagonal long-range correlation is associated with that of 
the ferromagnetic order whose origin is the strong correlation under the orbital degeneracy. 
Furthermore, we point out that $\eta$-pairing state emerges as the ground-state
in the present model in contrast to the single-orbital Hubbard Hamiltonian ~\cite{Yang:1989}, in which 
the state is not the ground-state but an excited eigen-state. 


\subsection{Trapped System}
\label{subsec:trap}

Next, we investigate the effects of the harmonic trap potential since 
typical atomic-gas experiments utilize the trap potential to 
prevent the escape of the atoms. 
We add the harmonic trap 
\mbox{$
H_{\rm trap} = \sum_{p,\sigma,i}V_{\rm ho}(i)n_{p,\sigma,i}
$ }
with \mbox{$V_{\rm ho}(i)=V[2/(L-1)]^2 [i-(L+1)/2]^2$} 
into Eq.~(\ref{eq:p1d}). 
Hereafter, we concentrate on highly imbalanced cases since
large $\tilde{m}$ favors the occurrence of the $\eta$-pairing state 
[Fig.~\ref{Fig2}] in the uniform system.   
In all the calculations, the harmonic-trap height is fixed as 
$V / t = 1.7$, and the total number of the spatial sites, $L=80$.  
We calculate the particle density
$n_{p}(i)=n_{p,\uparrow,i}+n_{p,\downarrow,i}$, the polarization density 
$m_{p}(i)=n_{p,\uparrow,i}-n_{p,\downarrow,i}$, and the particle density for each spin component,
$n_{p,\sigma,i}$. 
In addition, we examine the two-point pair correlation function, whose
base point is the trap center $L/2=40$, i.e., $C_{\rm pair}(i,40)$ with $i=40+r$ and $r \ge 1$.  
\begin{figure}[h]
\begin{center}
\includegraphics[width=1.0\linewidth]{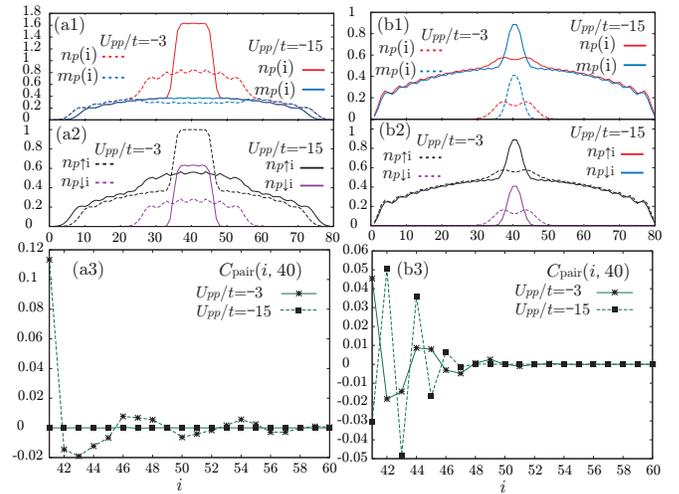}
\end{center}
\caption{(Color online) Ground-state properties in a trapped $p$-orbital
 chain, with (a) $N_{\uparrow}-N_{\downarrow}=40$ and (b)
 $N_{\uparrow}-N_{\downarrow}=60$.  
The total particle number is fixed as $N_{\uparrow}+N_{\downarrow}=68$. 
The trap potential is harmonic trap $H_{\rm trap}$, with potential 
strength $V/t=1.7$. 
The total number of the sites is $L=80$. 
The particle density $n_{p}(i)$ and the spin density $m_{p}(i)$ are
 shown in (a1,b1). 
The spin components $n_{p,\sigma,i}$ are also shown in (a2,b2).
The dashed and solid lines denote the results with weak ($U_{pp}/t =-3$) and 
strong ($U_{pp}/t =-15$) on-site interaction.
The two-point pair correlation functions, whose base point is the trap center
 $L/2=40$, are shown in (a3,b3). 
The filled box ($\blacksquare$) is for weak attractive
 interaction, whereas the asterisk symbol ($\ast$) is
 for strong attractive interaction.}
\label{Fig3}
\end{figure}


Figures \ref{Fig3}(a1,\,a2) show spatial profiles for
$n_{p}(i)$, $m_{p}(i)$ and $n_{p,\sigma,i}$ at
$N_{\uparrow}-N_{\downarrow}=40$. 
In addition, those for $N_{\uparrow}-N_{\downarrow}=60$ are 
shown in Fig.~\ref{Fig3}(b1,\,b2). 
Here, we focus on Figs.~\ref{Fig3}(a1) and (b1). 
It is found that the fully-spin-polarized region [i.e., $n_{p}(i)=m_{p}(i)$] 
appears at the trap edges.  
In addition, the particle density has a hump at the trap center.  
Comparing the results in $U_{pp}/t=-15$ (solid lines) with those in
$U_{pp}/t=-3$ (dashed lines), we find that for the large value of
$|U_{pp}|/t$, the fully-spin-polarized region enlarges whereas the area
of the particle-density hump shrinks. 
These behaviors are also checked by Figs.~\ref{Fig3}(a2) and (b2). 

The appearance of the fully-polarized regions is
insightful when exploring ordered states in trapped systems.  
Let us explain the origin of the fully-spin-polarized states in terms
of the attractive-repulsive transformation. 
The effect of the trap potential is regarded as a spatially-dependent
chemical potential, $\mu(i)=\mu-V_{\rm ho}(i)$. 
The variation is equivalent with application of a
spatially-modulating magnetic field $-\mu(i)\bar{S}_{p,i}^{z}$ into the
system. 
We find that $\mu(i)$ takes a negative value around the trap edges. 
When $|U_{pp}|/t$ and $N_\uparrow-N_\downarrow$ increase, the
ferromagnetic property of the system becomes predominant. 
The trap potential tends to store a polarized component
$\bar{S}_{p,i}^{z}\,(=\rho_{p,i}^{z}) <0$ on the trap edges. 
In other words, the transverse components of
the pseudo-spin-$1/2$ operator $\rho_{p,i}^{(\pm)}$ can vanish there.  
Since $N_{\uparrow}-N_{\downarrow} >0$, the up-spin components highly
concentrate at the trap edges. 
Thus, the superconducting pair can exist only in the density-profile
humps. 

We stress that the emergence of superconducting order at the trap center
depends on $|U_{pp}|/t$ and $N_{\uparrow}-N_{\downarrow}$. 
Here, let us check this point from the density profiles before examining
the pair-correlation function. 
Figures \ref{Fig3}(a1,\,a2) show that the hump around the trap
center for $U_{pp}/t=-15$ is regarded as another fully spin-polarized
region. 
The up-spin components fully occupy the sites at the trap center, since
$n_{p,\uparrow,i}=1$. 
It indicates that the down-spin components behaves as spinless
fermion. 
This local phase can be attributed to the PS phase seen in the uniform system. 
Although the polarized components on the trap-edge regions are 
detectable in single-band attractive Fermi gases~\cite{Tezuka;Ueda:2008}, 
the phase separation in the present system is more vivid (see Appendix B).

Now, let us examine the two-point pair correlation function 
around the central area in the trap potential. 
When $N_\uparrow-N_\downarrow=40$ and $U_{pp}/t=-3$, we find a spatially
oscillating pair correlation. 
The oscillation of the pair correlation indicates the emergence of the
FFLO superfluid phase in trapped system~\cite{Tezuka;Ueda:2008}. 
The amplitude of the two-point pair correlation completely vanishes in  
$N_\uparrow-N_\downarrow=40$ and $U_{pp}/t=-15$. 
This result is reasonable, since the motion of the up-spin component is
frozen as $n_{p,\uparrow,i}=1$. 
In contrast, when $N_\uparrow-N_\downarrow=60$ and $U_{pp}/t=-15$\,, 
the spin degrees of freedom on the trap central area survive 
[$n_{p,\uparrow,i}\neq 1$ in Fig.~\ref{Fig3}(b2)]. 
The pair correlation function has a staggered oscillating behavior for
$U_{pp}/t=-15$ as seen in Fig.~\ref{Fig3}(b3). 
The behavior is not obtained for $U_{pp}/t=-3$. 
We evaluate the particle density and the spin density at the trap
center. These densities exhibit $n_{p}(i=40) \simeq 1.24$ and $m_{p}(i=40)\simeq 0.47$. 
Since $\tilde{n}=2-n_{p}(i=40)\simeq 0.76$ and $\tilde{m}\simeq 0.47$,
the region allows to grow the $\eta$-paring phase as seen in Fig.~\ref{Fig2}(b). 
Thus, we propose that the $\eta$-paring phase occurs in the region 
surrounded by the polarized components in the presence of the trap potential. 
The $\eta$-pairing phase can be experimentally verified
by the detection scheme of the FFLO superfluid phase in trapped 
ultra-cold atomic gases.  
The measurement of the density profiles  such as phase-contrast polarization
imaging~\cite{Liao;Mueller:2010}, a time-of-flight method~\cite{Lu;Pu:2012}, and an interferometic
approach~\cite{Swanson;Trivedi:2012} allows us to obtain the information on
the wave vector $\bar{k}$ of the two-point pair correlation function.

\section{Summary}
\label{sec:summary}
In conclusion, we studied the pair correlation function in the 1-D $p$-orbital Fermi optical-lattice
with the tune of the attractive interaction and population imbalance.  
The DMRG calculations in the uniform system revealed that 
the off-diagonal long-range order characterized by $\eta$-pairing 
emerges for a large 
interaction strength and population imbalance range 
in addition to the FFLO superfluid and PS phases. 
Furthermore, we examined the effects of the harmonic trap potential. 
The $\eta$-paring phase clearly appears in a central region surrounded by
the spin-polarized phase. 

\begin{acknowledgments}
The numerical work was partially performed on Fujitsu BX900 in JAEA.   
This work was partially supported by the Strategic Programs for
Innovative Research, MEXT, and the Computational Materials Science
Initiative (CMSI), Japan. 
\end{acknowledgments}

\appendix

\section{multi-orbitals Hubbard Hamiltonian}
We start with the following Hamiltonian
\be 
H
&=&
\sum_{\sigma=\uparrow,\downarrow}\int d\bo{x}
\left[
\psi_{\sigma}^{\dagger}h_{0}\psi_{\sigma}
+
\frac{g}{2}
\psi_{\sigma}^{\dagger}\psi_{\bar{\sigma}}^{\dagger}
\psi_{\bar{\sigma}}\psi_{\sigma}
\right]\,, \lb{eq:hami}
\ee
with 
$
h_{0}
=
(-\hbar^2/2m)\nabla^2
+V_{\rm ver}+V_{\rm opt} 
$ and the coupling constant of the two-body interaction $g$. 
$V_{\rm ver}$ and $V_{\rm opt}$ are 
the cylindrically-symmetric vertical trap (on $xy$-plane) and the optical lattices
potential (along $z$-axis). 
The 1D $p$-orbital Hubbard Hamiltonian Eq.(\ref{eq:p1d}) is derived from Eq.(\ref{eq:hami}) using the
expansion 
$
\psi_{\sigma}=\sum_{\alpha}\sum_{i}
c_{\alpha,\sigma,i}u_{\alpha}w_{i}\,,
$
where $u_{\alpha}$ and $w_{i}$ are a wavefunction
associated with the eigensystem of 
\(
\left[(-\hbar^2/2m)\nabla_{\bot}^2+V_{\rm
ver}\right]u_{\alpha}=\epsilon_{\alpha}u_{\alpha}
\) and a Wannier function formed by the optical lattices. 
The indices $\alpha$ represent discrete energy levels caused by the
trap potential $V_{\rm ver}$. 
Now let us consider the situation where lowest energy level are completely occupied,
and  the $2$nd levels are partially filled. 
Then, including the $2$nd level corresponding to $\alpha=p_{x},p_{y}$-orbital and
taking the tight-binding approximation, we obtain the 1D $p$-orbital Hubbard Hamiltonian Eq.(\ref{eq:p1d}) with 
the hopping term 
$t=-\int dz\,w_{i+1}
\left(\frac{-\hbar^2}{2m}\frac{\partial^2}{\partial z^2}+V_{\rm opt}\right)
w_{i}$ 
and the on-site interaction 
$U_{\alpha\alpha'}=g\int d\bo{x}w_{i}^{4}u_{\alpha}^{2}
u_{\alpha'}^{2}$\,. 
Here, we assume that the vertical trap $V_{\rm ver}(\bo{x}_{\bot})$ is harmonic trap potential. 
Then, using the exact solution of the two dimensional harmonic oscillator, 
the relation between intra-orbital interaction $U_{pp}$ and inter-band interaction $U_{p_{x}p_{y}}$ is 
given by $U_{p_{x}p_{y}}=U_{pp}/3$\,.

\section{Trapped single orbital 1D attractive Hubbard chain}
We show DMRG calculation results for single-band Hubbard model 
with harmonic trap potential,
\be
&&H
=
-t
\sum_{i,\sigma}
(
c_{\sigma,i+1}^{\dagger}c_{\sigma,i}
+
\text{h.c.}
) +U\sum_{i}n_{\uparrow,i}n_{\downarrow,i}\nn \\
&&\qquad+\sum_{\sigma,i}V\left(
\frac{i-(L+1)/2 }{(L-1)/2}
\right)^2
n_{\sigma,i}
\,, 
\ee
\begin{figure}[h]
\begin{center}
\includegraphics[width=1.0\linewidth]{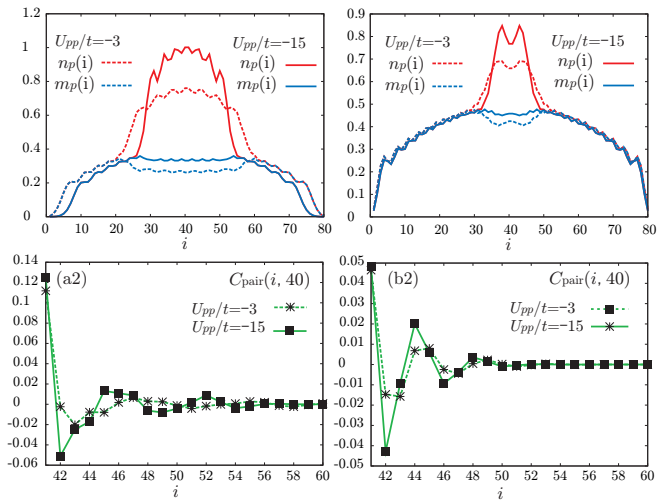}
\end{center}
\caption{(Color online) 
Ground-state properties in a trapped single orbital 1D attractive Hubbard chain 
with (a) $N_{\uparrow}-N_{\downarrow}=20$ and  (b) $N_{\uparrow}-N_{\downarrow}=30$\,. 
The total particle number is fixed as $N_{\uparrow}+N_{\downarrow}=34$. 
The results with these population imbalance are comparable to 
Fig.~\ref{Fig3} in main article. 
The particle density $n(i)=n_{\uparrow,i}+n_{\downarrow,i}$ and the spin density $m(i)=n_{\uparrow,i}+n_{\downarrow,i}$ are shown in (a1,b1)\,. 
Two-point pair correlation functions are shown in (a2,b2)\,. 
The other parameters are $U/t=-3 ,-15$\,, $V/t=1.7$ and $L=80$.
}
\label{Fig4}
\end{figure}

Fig~\ref{Fig4} show the (spin-)density profiles and the two-point pair correlation functions. 
The population imbalance is set to be $N_{\uparrow}-N_{\downarrow}=20$ and 
$N_{\uparrow}-N_{\downarrow}=30$ via the total particle number $N_{\uparrow}+N_{\downarrow}=36$\,. 
Then, population imbalance rates $P\equiv(N_{\uparrow}-N_{\downarrow})/(N_{\uparrow}+N_{\downarrow})$ 
of Fig~\ref{Fig4} become same with that of Fig.\ref{Fig3} in main text 
((a) $P=5/9$ and (b) $P=5/6$). 
We find that the results with small attractive interaction $U/t=-3$ are very similar with that of $p$-orbital 1D chain (see Fig.\ref{Fig3} in main article). 
Increasing $|U|/t$, the polarized regions $(n(i)=m(i))$ are enlarged as $p$-orbital chain. 
But the enlargement of the polarized regions is small in comparison with Fig.\ref{Fig3}(a1,b1) in main text. 
The vanishment behavior and staggered oscillation of pair correlation functions as Fig.\ref{Fig3}(a3,b3) are not observed 
in single band Hubbard chain. 
The behavior of the (spin-)density profiles and correlation functions for large attractive interaction $U/t=-15$ are qualitative different with $p$-orbital 1D chain.

\end{document}